\documentclass[12pt]{article}

\usepackage{graphics}
\usepackage{graphicx,epsfig}
\usepackage{epsfig}
\usepackage{graphicx}
\usepackage{a4wide}
\usepackage{amsmath, amssymb}
\usepackage{amsfonts}
\usepackage{enumerate}

\newcommand{\pf}{{\bf Proof : }}

\newtheorem{theorem}{Theorem}[section]

\newtheorem{proposition}{Proposition}[section]

\newtheorem{corollary}{Corollary}[section]
\newtheorem{example}{Example}[section]

\newcommand{\Z}{\mathbb{Z}}

\newcommand{\R}{\mathbb{R}}

\newcommand{\xn}{x^n - 1}

\begin{document}

\title{\bf 
A class of cyclic Codes Over the Ring $\Z_4[u]/<u^2>$ and its Gray image
}
\author{Sukhamoy Pattanayak and Abhay Kumar Singh\\
Department of Applied Mathematics\\ 
         Indian School of Mines\\
         Dhanbad 826 004,  India \\
e-mail : sukhamoy88@gmail.com \\ singh.ak.am@ismdhanbad.ac.in \\
}
\date{}
\maketitle

\begin{abstract}
Cyclic codes over R have been introduced recently. In this paper, we study the cyclic codes over R and their $\Z_2$ image. Making use of algebraic structure, we find the some good $\Z_2$ codes of length 28.
\end{abstract}

\section{Introduction}
The study of cyclic codes over finite ring is a topic of growing interest due to their rich algebraic structure. In the seminal paper [3], Hammons et al. established connection between some good binary nonlinear codes and $\Z_4$-linaer codes via gray map. The different aspect of cyclic codes over $\Z_4$ and some other finite rings have been discussed in series of papers[1,11-13]. In[15], Yildiz and Aydin discussed the cyclic codes over $\Z_4+u\Z_4$ and their $\Z_4$ image and added some good codes to the data base of $\Z_4$ codes. The cyclic codes of odd length over $\Z_4+u\Z_4$ are studied in[10]. Yildiz and karadeniz discussed the linear codes over $\Z_4+u\Z_4$ in[2].\\
The ring  $\Z_4+u\Z_4$ is extension of $\Z_4$ and it is a frobenius ring. We extended the result of Yildiz and Aydin by studying the $\Z_2$ image of  $\Z_4+u\Z_4$. The class of quasi-cyclic codes(QC) is a great platform of obtaining good codes which attain a version of Gilbert Varshamov bound. These finding motivated the study of cyclic codes over  $\Z_4+u\Z_4$ and their $\Z_2$ gray image, because as we can directly be shown, the $\Z_2$ images are equivalent to 4-QC codes over $\Z_2$. With the help of example, we find some good $\Z_2$ codes through cyclic codes over  $\Z_4+u\Z_4$.\\
This paper is organized as follows. In section 2, we have given some preliminaries related with this work. In sec 3, we find gray image of cyclic codes over $\Z_4+u\Z_4$. We have reviewed some results of cyclic codes of $\Z_4+u\Z_4$ in sec 4. In sec 5, we have provided some good $\Z_2$ codes of length 28.

\section{Preliminaries}
Let $R$ be the commutative,characteristic 4 ring $\Z_4+u\Z_4=\{a+ub\vert a,b \in \Z_4 \}$ with $u^2=0$. $R$ can also be thought of as the quotient ring $ \Z_4[u]/\langle u^2 \rangle$. The units of $R$ are
\begin{center}
$1, 3, 1 + u, 1 + 2u, 1 + 3u, 3 + u, 3 + 2u, 3 + 3u$,
\end{center}
and the non-units are
\begin{center}
$0, 2, u, 2u, 2+ u, 2 + 2u, 3u, 2 + 3u$.
\end{center}
R has six ideals in all listed below:
\begin{center}
$\{0\},\langle u \rangle,\langle 2 \rangle,\langle 2u \rangle,\langle 2+u \rangle,\langle 2,u \rangle$.
\end{center}
$R$ is a non-principal local ring with $\langle 2,u \rangle$ as its unique maximal ideal. A commutative ring is called a chain ring if its ideals form a chain under the relation of inclusion. But from the above we see that the ideals of $R$ do not form chain. Therefore, $R$ is a non-chain ring. \\
A linear code $C$ of length $n$ over $R$ is a $R$-submodule of $R^n$. An element of $C$ is called a codeword.
A code of length $n$ is cyclic if the code is invariant under the automorphism $\sigma$ which has
\begin{center}
$\sigma(c_0, c_1, \cdots , c_{n-1}) = (c_{n-1}, c_0, \cdots , c_{n-2}).$
\end{center}
A code of length $n$ is 4-quasi cyclic if the code is invariant under the automorphism $\nu$ which has
\begin{center}
$\nu(c_0, c_1, \cdots ,c_{n-4},c_{n-3},c_{n-2},c_{n-1}) = (c_{n-4},c_{n-3},c_{n-2},c_{n-1}, c_0, \cdots , c_{n-5}).$
\end{center}
It is well known that a cyclic code of length $n$ over $R$ can be identified with an ideal in the quotient ring $R[x]/\langle x^n-1\rangle$ via the $R$-module isomorphism
as follows:
\begin{center}
$R^n \longrightarrow R[x]/\langle x^n-1\rangle$\\
$(c_0,c_1,\cdots,c_{n-1}) \mapsto c_0+c_1x+\cdots+c_{n-1}x^{n-1} (\text{mod}\langle x^n-1\rangle) $
\end{center}
The residue field is given by $R/\langle 2,u \rangle \cong F_2$. The image of any element $a \in R$ under the projection map $\mu: R \longrightarrow \overline{R}$ is denoted by $\overline{a}$. The map $\mu$ is extended to $R[x] \longrightarrow \overline{R}[x]$ in the natural way. A polynomial $f(x) \in R[x]$ is called basic irreducible (primitive) if $f(x)$ is an irreducible (primitive) polynomial in $R[x]$. A polynomial $f(x)$ over $R$ is called a regular polynomial if it is not a zero divisor in $R[x]$, equivalently, $f(x)$ is regular if $f(x)\neq 0$. Two polynomials $f(x), g(x) \in R[x]$ are said to be coprime if there exist $a(x), b(x) \in R[x]$ such that $a(x)f(x) + b(x)g(x) = 1$.
\begin{theorem}(Hensel's Lemma{[13]})
Let $f$ be a monic polynomial in $\Z_4[x]$ and assume that $f (\text{mod} ~2) = g_1g_2 \cdots g_r$, where $g_1, g_2,\cdots , g_r$ are pairwise coprime monic polynomials over $F_2$. Then there exist pairwise coprime monic polynomials $f_1, f_2,\cdots , f_r$ over $\Z_4$ such that $f = f_1f_2 \cdots f_r$ in $\Z_4[x]$ and $f_i (\text{mod} ~2) = g_i, i = 1, 2, \cdots , r$.
\end{theorem}

\section{Gray images of cyclic codes over $\Z_4+u\Z_4$}
The Lee weight was defined as $w_L(a)= min \lbrace a,4-a \rbrace, a\in \Z_4$ i.e, $w_L(0)=0,w_L(1)=1,w_L(2)=2,w_L(3)=1$. Let $a+ub$ be any element of the ring $R=\Z_p+u\Z_p, u^2=0$. The Lee Weight $w_L$ of the ring $R$ is defined as follows 
\begin{center}
$w_L(a+ub)=w_L((b,a+b))$,
\end{center}
where $w_L((b,a+b))$ described the usual Lee weight on $\Z_4^2$. For any $c_1,c_2 \in R$, the Lee distance $d_L$, given by $d_L(c_1,c_2)=w_l(c_1-c_2)$. The minimum Lee distance of $C$ is the smallest nonzero Lee distance between all pairs of distinct codewords of $C$. The Hamming weight of a codeword $c=(c_0,c_1,\cdots,c_{n-1})$ denoted by $w_H(c)$ is the number of non zero entries in $c$. The Hamming distance $d(c_1, c_2)$ between two codewords $c_1$ and $c_2$ is the Hamming weight of the codeword $c_1-c_2$. The minimum Hamming distance of a linear code $C$ is given by
\begin{center}
$d_H(C)= min \lbrace d(c_1, c_2): c_1,c_2 \in C, c_1 \neq c_2  \rbrace$.
\end{center}
Now we define the Gray map on $R$. Any element $c \in R$ can be expressed as $c=a+ub$, where $a,b \in \Z_4$. The Gray map defined as follows 
\begin{center}
$\phi:\Z_4 + u \Z_4 \longrightarrow \Z_4^2$\\
such that $\phi (a+ub)=(b,a+b)$~~~~~~~~$a,b \in Z_4$
\end{center}
Again we give the definition of the Gray map from $\Z_4$ to $\Z_2^2$. First we see that the 2-adic expansion of $c\in \Z_4$ is $c=\alpha(c)+2\beta(c)$ such that $\alpha(c)+\beta(c)+\gamma(c)=0$ for all $c\in \Z_4$. \\
Then we get the table below
\begin{center}
\begin{tabular}{ l  c  c  c }

$c$&$\alpha(c)$&$\beta(c)$&$\gamma(c)$ \\
$0$&$0$&$0$&$0$ \\
$1$&$1$&$0$&$1$ \\
$2$&$0$&$1$&$1$ \\
$3$&$1$&$1$&$0$ \\

\end{tabular}
\end{center} 
The Gray map ~~~ $\psi:\Z_4 \longrightarrow \Z_2^2$ ~given by~ $\psi(c)=(\beta(c),\gamma(c))$,~~$c \in \Z_4$ i.e, $\psi(0)=(0,0),\psi(1)=(0,1),\psi(2)=(1,1),\psi(3)=(1,0)$. Then we get the composite map as follows
\begin{center}
$\Phi(\psi \cdot \phi):R\longrightarrow \Z_2^4$\\
$\Phi(a+ub)=\psi(\phi(a+ub))=\psi(b,a+b)$\\
$~~~~~~~~~~~~~~=(\beta(b),\gamma(b),\beta(a+b),\gamma(a+b))$.
\end{center}
It is well known that $\Phi$ is a linear map on $R$ and is an isometry from($R$, Lee distance)
to($\Z_2^4$ , Hamming distance). This map $\Phi$ can be extended to $R^n$ in the natural way:\\~\\ 
$\Phi:R^n\longrightarrow \Z_2^4n$\\
$(c_0,c_1,\cdots,c_{n-1})\longrightarrow (\beta(b_0),\gamma(b_0),\beta(a_0+b_0),\gamma(a_0+b_0),\beta(b_1),\gamma(b_1),\beta(a_1+b_1),\\ ~~~~~~~~~~~~~~~~~~~~~~~~~~~~~\gamma(a_1+b_1),\cdots,\beta(b_{n-1}),\gamma(b_{n-1}), \beta(a_{n-1}+b_{n-1}),\gamma(a_{n-1}+b_{n-1}))$\\
where $c_i = a_i + ub_i, 0\leq i\leq n-1$.
\begin{proposition}
The Gray map $\Phi$ is a distance-preserving map from ($R^n$, Lee distance) to ($\Z_2^{4n}$ , Hamming distance) and this map also $\Z_2$ linear.
\end{proposition}
\pf
From the definitions, it is clear that $\Phi(c_1-c_2)=\Phi(c_1)-\Phi(c_2)$ for $c_1,c_2 \in R^n$. Thus, $d_L(c_1,c_2)=w_L(c_1-c_2)=w_H(\Phi(c_1-c_2))=w_H(\Phi(c_1)-\Phi(c_2))=d_H(\Phi(c_1),\Phi(c_2))$. Let $c_1,c_2 \in R^n, k_1,k_2 \in \Z_2$, then from the definition of the Gray map, we have\\ $\Phi(k_1c_1+k_2c_2)=k_1\Phi(c_1)+k_2\Phi(c_2)$, that implies $\Phi$ is $\Z_2$ linear.
\begin{theorem}
Let $\sigma$ be the cyclic shift of $R^n$ and $\nu$ denote the 4-QC shift of $\Z_2^{4n}$. Let $\Phi$ be the Gray map from $R^n$ to $\Z_2^{4n}$. Then prove that $\Phi \sigma=\nu \Phi$
\end{theorem}
\pf
Let $c=(c_0,c_1,\cdots,c_{n-1})\in R^n$, where $c_i = a_i + ub_i$,with $a_i,b_i \in \Z_4, 0\leq i\leq n-1$. From the definition of the Gray map, we get \\
$\Phi(c) = (\beta(b_0),\gamma(b_0),\beta(a_0+b_0),\gamma(a_0+b_0),\beta(b_1),\gamma(b_1),\beta(a_1+b_1),\gamma(a_1+b_1),\cdots,\beta(b_{n-1}), \\ \gamma(b_{n-1}), \beta(a_{n-1}+b_{n-1}),\gamma(a_{n-1}+b_{n-1}))$\\
Hence \\
$\nu (\Phi(c))=(\beta(b_{n-1}),\gamma(b_{n-1}), \beta(a_{n-1}+b_{n-1}),\gamma(a_{n-1}+b_{n-1}),\beta(b_0),\gamma(b_0),\beta(a_0+b_0),\gamma(a_0+b_0),\cdots,\beta(b_{n-2}),\gamma(b_{n-2}), \beta(a_{n-2}+b_{n-2}),\gamma(a_{n-2}+b_{n-2}))$. \\
On the other hand,\\ 
$\sigma(c)=(c_{n-1},c_0,c_1,\cdots,c_{n-2})$\\
we deduce that \\
$\Phi(\sigma(c))=(\beta(b_{n-1}),\gamma(b_{n-1}), \beta(a_{n-1}+b_{n-1}),\gamma(a_{n-1}+b_{n-1}),\beta(b_0),\gamma(b_0),\beta(a_0+b_0),\gamma(a_0+b_0),\cdots,\beta(b_{n-2}),\gamma(b_{n-2}), \beta(a_{n-2}+b_{n-2}),\gamma(a_{n-2}+b_{n-2}))$. \\
Therefore ~~~$\Phi \sigma=\nu \Phi$.
\begin{theorem}
A linear code $C$ of length $n$ over $R$ is a cyclic code if and only if $\Phi(C)$ is a 4-quasi cyclic code of length $4n$ over $\Z_2$.
\end{theorem}
\pf
It is immediately get from previous theorem.
\begin{corollary}
The Gray image of a cyclic code of length $n$ over $R$ is a distance invariant linear 4-quasi cyclic code of length $4n$ over $\Z_2$.
\end{corollary}
\pf
Let $C$ be a  cyclic code of length $n$ over $R$. Then $\sigma(C)=C$, therefore $\Phi(\sigma(C))=\Phi(C)$. It follows from theorem 3.1 that $\nu(\Phi(C))=\Phi(C)$, which means that $\Phi(C)$ is a 4-quasi cyclic code.

\section{Cyclic codes}
Let length $n$ is odd through out that section. For a finite chain ring $\R$, it is well known that the ring $\frac{\R[x]}{\langle x^n-1\rangle}$ is a principal ideal ring. But in that case the ring $R=\Z_4+u\Z_4$,$u^2=0$ is not a chain ring and the situation is not so easy. 
\begin{proposition}([10])
The ring $R_n = \frac{R[x]}{\langle x^n-1\rangle}$ is not a principal ideal ring.
\end{proposition}
Therefore, a cyclic code of length $n$ over $R$ is not principally generated. As $n$ is odd, the ring $\frac{\Z_4[x]}{\langle x^n-1\rangle}$ is a principal ideal ring. So a cyclic code of length $n$ over $R$ is of the form $C = C_1 +uC_2 = \langle g_1\rangle+u\langle g_2\rangle$, where $g_1, g_2 \in \Z_4[x]$ are generator polynomials of the cyclic codes $C_1, C_2$, respectively.\\
We assume the ideals of $R[x]/ \langle g\rangle$, where $g$ is a basic irreducible polynomial over $R$.
\begin{theorem}
If $g \in R[x]$ be a basic irreducible polynomial. Then the ideals of
$R[x]/ \langle g\rangle$ are precisely, $\lbrace 0\rbrace, \langle 1 + \langle g\rangle\rangle, \langle 2 + \langle g\rangle\rangle, \langle u + \langle g\rangle\rangle, \langle 2u + \langle g\rangle\rangle, \langle 2+u + \langle g\rangle\rangle$ and $\langle\langle 2,u\rangle + \langle g\rangle\rangle$.
\end{theorem}
\begin{theorem}
Let $x^n-1 = g_1g_2 \cdots g_m$, where $g_i, i = 1, 2,\cdots,m$ are basic irreducible pairwise coprime polynomials in $R[x]$. Then any ideal in $R_n$ is the sum of the ideals of $R[x]/ \langle g_i\rangle, i = 1, 2,\cdots,m$.
\end{theorem}
\pf
It follows from the Chinese Remainder Theorem.
\begin{corollary}
There are $7^m$ cyclic codes of length $n$ over $R$.
\end{corollary}
\begin{theorem}
A linear code $C = C_1 +uC_2$ of length $n$ over $R$ is cyclic if and only if $C_1, C_2$ are cyclic codes of length $n$ over $\Z_4$.
\end{theorem}
Define $\psi : R \longrightarrow \Z_4 $ by $\psi(a + ub) = a ~\text{(mod u)}~,$ where $a, b \in \Z_4$. The map $\psi$ is a ring homomorphism. We extend the map $\psi$ to a homomorphism $\phi : \frac{R[x]}{\langle x^n-1\rangle} \longrightarrow \frac{\Z_4[x]}{\langle x^n-1\rangle}$ defined by
$$ \phi(c_0 + c_1x + \cdots + c_{n-1}x^{n-1}) = \psi(c_0) + \psi(c_1)x + \cdots + \psi(c_{n-1})x^{n-1},$$ 
where $c_i \in R$. We have $\text{ker} \phi = \langle u \rangle = u\Z_4$. \\
Let $C$ be a cyclic code of length $n$ over $R$. Restrict $\phi$ to $C$ and define
$$ J=\lbrace h(x)\in \frac{\Z_4[x]}{\langle x^n-1\rangle} : uh(x)\in  \text{ker} \phi \rbrace. $$
Obviously $J$ is an ideal of $\frac{\Z_4[x]}{\langle x^n-1\rangle}$. So $J$ is a cyclic code over $\Z_4$. Again, the image of $C$ under $\phi$ is an ideal of  $\frac{\Z_4[x]}{\langle x^n-1\rangle}$.\\
Let $n$ be an odd integer. Then the cyclic code of length $n$ over $\Z_4$ is principally generated. So $\phi(C)$ and $\text{ker} \phi$ are principal ideals of $\frac{\Z_4[x]}{\langle x^n-1\rangle}$, and $\phi(C)=\langle f_1(x)+2f_2(x)\rangle$ and $\text{ker} \phi=\langle uf_3(x)+2uf_4(x)\rangle$, where $f_2(x)\vert f_1(x)\vert \xn$ and $f_4(x)\vert f_3(x)\vert \xn$. Therefore $C=\langle f_1(x)+2f_2(x)+uf_{13}(x)+2uf_{14}(x),uf_3(x)+2uf_4(x) \rangle$. Using the theorem which is used in [2], we get the following result.
\begin{theorem}
Let $n$ be an odd integer and $C$ be a cyclic code of length $n$ over $R$. Then
$$ C=\langle f_1(x)+2f_2(x)+2uf_{14}(x),uf_3(x)+2uf_4(x) \rangle ,$$
where $f_2(x)\vert f_1(x)\vert \xn$ and $f_4(x)\vert f_3(x)\vert \xn$ in $\frac{R[x]}{\langle x^n-1\rangle}$. 
\end{theorem}
We Know that the basis of $C$ over $R$ is called the minimal generating set of $C$, and the number of element include in the minimal generating set is called the rank of the code $C$, denoted the $rank(C)$.
\begin{theorem}
Let $C$ be a cyclic code of length $n$ over $R$. If $C =\langle f_1(x)+2f_2(x)+2uf_{14}(x),uf_3(x)+2uf_4(x)\rangle$ and $deg f_1(x)=k_1$ and $deg f_2(x)=k_2$, then $C$ has rank $n-k_2$ and a minimal spanning set $A = \{ (f_1(x)+2f_2(x)+2uf_{14}(x)), x(f_1(x)+2f_2(x)+2uf_{14}(x)), \dots, x^{n-k_1-1}(f_1(x)+2f_2(x)+2uf_{14}(x)), u(f_3(x)+2f_4(x)), xu(f_3(x)+2f_4(x)), \dots,\\ x^{k_1-k_2-1}u(f_3(x)+2f_4(x))\}$.
\end{theorem}

\section{Example}
In this section, we give some examples of cyclic codes of different lengths over the ring $R$.

\begin{example}
Cyclic codes of length $3$ over $R = \Z_4 + u \Z_4 , u^2 = 0$: We have
$$ x^3-1 = (x-1)(x^2 + x +1) ~\text{over}~ R.$$ 
Let $g_1 = x-1$ and $g_2=x^2 + x + 1$. The cyclic codes of length $3$ over $R$ are given in Table 1.
\end{example}
\begin{center}
{\bf Table 1.} Non-zero cyclic codes of length 3 over $\Z_4 + u \Z_4, u^2 = 0$.\\~\\
\begin{tabular}{| l | c |}
\hline
 Non-zero generator polynomials & ranks  \\
\hline
$<2g_i,ug_1+2u>, i=1,2.$ & 2  \\
\hline
$<2g_i,ug_2+2u>, i=1,2.$ & 1 \\
\hline
$<2g_i,3u>, i=1,2.$ & 3 \\
\hline
$<2,ug_1+2u>$ & 2 \\
\hline
$<2,ug_2+2u>$ & 1 \\
\hline
$<2,3u>$ & 3 \\
\hline
$<g_1+2,3u>$ & 3 \\
\hline
$<g_2+2,3u>$ & 3 \\
\hline
\end{tabular}
\end{center}

\begin{example}
Cyclic codes of length $7$ over $R = \Z_4 + u \Z_4 , u^2 = 0$: We have
$$ x^7-1 = (x-1)(x^3 + x +1)(x^3 + x^2 + 1) ~\text{over}~ F_2.$$ 
This factors are irreducible polynomials over $F_2$. The Hensel lifts of $x^3 + x +1$ to $\Z_4$ is $x^3 + 2x^2 + x - 1$ and Hensel lifts of $x^3 + x^2 +1$ to $\Z_4$ is $x^3 - x^2 - 2x - 1$. Therefore we have
$$ x^7-1 = (x-1)(x^3 + 2x^2 + x - 1)(x^3 - x^2 - 2x - 1) ~\text{over}~ R.$$
Let $g_1 = x-1$, $g_2=x^3 + 2x^2 + x - 1$ and $g_3 = x^3 - x^2 - 2x - 1.$ The cyclic codes of length $7$ over $R$ are given in Table 2.
\end{example}
\begin{center}
{\bf Table 2.} Non-zero cyclic codes of length 7 over $\Z_4 + u \Z_4, u^2 = 0$.\\~\\
\begin{tabular}{| l | c |}
\hline
 Non-zero generator polynomials &ranks \\
\hline
$<2g_ig_j, ug_1g_2+2ug_1>, i\neq j=1,2,3.$ & 3 \\
\hline
$<2g_ig_j, ug_1g_2+2ug_2>, i\neq j=1,2,3.$ & 3 \\
\hline
$<2g_ig_j, ug_1g_2+2u>, i\neq j=1,2,3.$ & 3 \\
\hline
$<2g_ig_j, ug_1g_3+2ug_1>, i\neq j=1,2,3.$ & 3  \\
\hline
$<2g_ig_j, ug_1g_3+2ug_3>, i\neq j=1,2,3.$ & 3 \\
\hline
$<2g_ig_j, ug_1g_3+2u>, i\neq j=1,2,3.$ & 3 \\
\hline
$<2g_ig_j, ug_2g_3+2ug_2>, i\neq j=1,2,3.$ & 1 \\
\hline
$<2g_ig_j, ug_2g_3+2ug_3>, i\neq j=1,2,3.$ & 1 \\
\hline
$<2g_ig_j, ug_2g_3+2u>, i\neq j=1,2,3.$ & 1 \\
\hline
$<2g_ig_j, ug_1+2u>, i\neq j=1,2,3.$ & 6 \\
\hline
$<2g_ig_j, ug_2+2u>, i\neq j=1,2,3.$ & 4 \\
\hline
$<2g_ig_j, ug_3+2u>, i\neq j=1,2,3.$ & 4 \\
\hline
$<2g_ig_j, 3u>, i\neq j=1,2,3.$ & 7 \\
\hline
$<2g_i, ug_1g_2+2ug_1>, i=1,2,3.$ & 3 \\
\hline
$<2g_i, ug_1g_2+2ug_2>, i=1,2,3.$ & 3 \\
\hline
$<2g_i, ug_1g_2+2u>, i=1,2,3.$ & 3 \\
\hline
$<2g_i, ug_1g_3+2ug_1>, i=1,2,3.$ & 3  \\
\hline
$<2g_i, ug_1g_3+2ug_3>, i=1,2,3.$ & 3 \\
\hline
$<2g_i, ug_1g_3+2u>, i=1,2,3.$ & 3 \\
\hline
$<2g_i, ug_2g_3+2ug_2>, i=1,2,3.$ & 1 \\
\hline
$<2g_i, ug_2g_3+2ug_3>, i=1,2,3.$ & 1 \\
\hline
$<2g_i, ug_2g_3+2u>, i=1,2,3.$ & 1 \\
\hline
$<2g_i, ug_1+2u>, i=1,2,3.$ & 6 \\
\hline
$<2g_i, ug_2+2u>, i=1,2,3.$ & 4 \\
\hline
$<2g_i, ug_3+2u>, i=1,2,3.$ & 4 \\
\hline
$<2g_i, 3u>, i=1,2,3.$ & 7 \\
\hline
$<2, ug_1g_2+2ug_1>.$ & 3 \\
\hline
$<2, ug_1g_2+2ug_2>.$ & 3 \\
\hline
$<2, ug_1g_2+2u>.$ & 3 \\
\hline
$<2, ug_1g_3+2ug_1>.$ & 3 \\
\hline
$<2, ug_1g_3+2ug_3>.$ & 3 \\
\hline
$<2, ug_1g_3+2u>.$ & 3 \\
\hline
$<2, ug_2g_3+2ug_2>.$ & 1 \\
\hline
$<2, ug_2g_3+2ug_3>.$ & 1 \\
\hline
$<2, ug_2g_3+2u>.$ & 1 \\
\hline
$<2, ug_1+2u>.$ & 6 \\
\hline
$<2, ug_2+2u>.$ & 4  \\
\hline
$<2, ug_3+2u>.$ & 4 \\
\hline
$<2, 3u>.$ & 7 \\
\hline
\end{tabular}
\end{center}
\begin{center}
{\bf Table 2.} Non-zero cyclic codes of length 7 over $\Z_4 + u \Z_4, u^2 = 0$.\\~\\
\begin{tabular}{| l | c |}
\hline
 Non-zero generator polynomials & ranks \\
\hline
$<g_1g_2+2g_1, ug_1+2u>.$ & 6 \\
\hline
$<g_1g_2+2g_1, ug_2+2u>.$ & 4 \\
\hline
$<g_1g_2+2g_1, 3u>.$ & 7 \\
\hline
$<g_1g_2+2g_2, ug_1+2u>.$ & 6 \\
\hline
$<g_1g_2+2g_2, ug_2+2u>.$ & 4 \\
\hline
$<g_1g_2+2g_2, 3u>.$ & 7 \\
\hline
$<g_1g_2+2, ug_1+2u>.$ & 6 \\
\hline
$<g_1g_2+2, ug_2+2u>.$ & 4 \\
\hline
$<g_1g_2+2, 3u>.$ & 7 \\
\hline

$<g_1g_3+2g_1, ug_1+2u>.$ & 6 \\
\hline
$<g_1g_3+2g_1, ug_3+2u>.$ & 4  \\
\hline
$<g_1g_3+2g_1, 3u>.$ & 7 \\
\hline
$<g_1g_3+2g_3, ug_1+2u>.$ & 6 \\
\hline
$<g_1g_3+2g_3, ug_3+2u>.$ & 4 \\
\hline
$<g_1g_3+2g_3, 3u>.$ & 7 \\
\hline
$<g_1g_3+2, ug_1+2u>.$ & 6 \\
\hline
$<g_1g_3+2, ug_3+2u>.$ & 4 \\
\hline
$<g_1g_3+2, 3u>.$ & 7 \\
\hline

$<g_2g_3+2g_2, ug_2+2u>.$ & 4 \\
\hline
$<g_2g_3+2g_2, ug_3+2u>.$ & 4 \\
\hline
$<g_2g_3+2g_2, 3u>.$ & 7 \\
\hline
$<g_2g_3+2g_3, ug_2+2u>.$ & 4 \\
\hline
$<g_2g_3+2g_3, ug_3+2u>.$ & 4 \\
\hline
$<g_2g_3+2g_3, 3u>.$ & 7 \\
\hline
$<g_2g_3+2, ug_2+2u>.$ & 4 \\
\hline
$<g_2g_3+2, ug_3+2u>.$ & 4 \\
\hline
$<g_2g_3+2, 3u>.$ & 7 \\
\hline
$<g_1+2, 3u>.$ & 7 \\
\hline
$<g_2+2, 3u>.$ & 7 \\
\hline
$<g_3+2, 3u>.$ & 7 \\
\hline
\end{tabular}
\end{center}

\end{document}